\documentclass[aps,prd,twocolumn,preprintnumbers,nofootinbib,showpacs]{revtex4}
\usepackage{amsmath,amsfonts,amssymb,bm}
\usepackage{graphicx}
\usepackage{color}
\usepackage{subfigure}
\usepackage{multirow}
\usepackage{textcomp}
\usepackage{slashed}

\definecolor{purple}{rgb}{0.5,0,0.5}
\definecolor{blue}{rgb}{0.0,0,0.9}

\usepackage[colorlinks=true, pdfstartview=FitV, linkcolor=purple, citecolor= purple, urlcolor=blue]{hyperref}

\begin{document}


\title{$B_c$ Meson Spectrum Via Dyson-Schwinger Equation and Bethe-Salpeter Equation Approach}

\author{Muyang Chen}\email{muyang@nankai.edu.cn}
\affiliation{Department of Physics, Hunan Normal University, Changsha 410081, China}

\author{Lei Chang}\email{leichang@nankai.edu.cn}
\affiliation{School of Physics, Nankai University, Tianjin 300071, China}

\author{Yu-xin Liu}\email{yxliu@pku.edu.cn}
\affiliation{Department of Physics and State Key Laboratory of Nuclear Physics and Technology, Peking University, Beijing 100871, China}
\affiliation{Collaborative Innovation Center of Quantum Matter, Beijing 100871, China}
\affiliation{Center for High Energy Physics, Peking University, Beijing 100871, China}

\date{\today}

\begin{abstract}
 We predict the masses of the lowlying $B_c$ mesons with $J^P = 0^-,\,1^-,\,0^+,\,1^+,\,2^+$, using a flavor dependent interaction pattern which gives an unified successful description of the light, heavy-light and heavy mesons and is also appliable to the radial excited heavy mesons.
 The errors are controlled carefully.
 With the errors from the RL approximation subduced, our predictions are consistent with the lQCD and quark model results, which supports strongly that the flavor dependent interaction pattern is reasonable. Our predictions provide significant guides to the experiment search of the $B_c$ mesons. 

\end{abstract}

\maketitle


\section{Introduction}

The $B_c$ mesons are the only heavy mesons containing two different flavor valence quarks, one charm quark and one bottom quark. The quark flavor difference forbids their annihilation into gluons or photons, so that the ground state pseudoscalar $B_c(1S)$ can only decay weakly, which makes it particularly interesting  for the study of weak interaction. The excited $B_c$ states lying below the BD production threshold decay into the ground state via hadronic or radiative transitions. As a result, the $B_c$ mesons are supposed to be much more stable than their counterparts in the $c\bar{c}$ and $b\bar{b}$ systems, and spin splittings can be studied in detail. 
 However, in the experimental aspect, the $B_c$ mesons are much less explored than the charmonium and bottomonium due to the small production rate, as the dominant production mechanism requires the production of both $c\bar{c}$ and $b\bar{b}$ pairs. The ground state pseudoscalar $B_c(1S)$ meson is first observed in 1998 by CDF collaboration at the Tevatron collider \cite{Abe1998}. 
The vector meson, $B_c^*(1S)$, decays into $B_c(1S)$ via radiative transition,
$B_c^*(1S) \to B_c(1S)\gamma$. It's challenging to detect the emitted photon due to the low energy. The vector meson mass, $M_{B_c^*(1S)}$, is still not determined. Until recently two excited $B_c$ states were reported by CMS collaboration \cite{Sirunyan2019} and LHCb collaboration \cite{Aaij2019}, pushing forward the study of the $B_c$ mesons. More $B_c$ mesons are expected to be discovered \cite{Chang2015}.
 In the theoretical aspect, the prediction of the $B_c$ mesons have existed since 40 years ago \cite{Eichten1981}, and vairous model calculations occured in these year \cite{Kwong1991,Eichten1994,Gershtein1995,Gupta1996,Fulcher1999,Ebert2003,Godfrey2004,Li2019}. However, these predictions vary widely. For example, the 1S-spin splitting, i.e. $M_{B^*_c(1S)} - M_{B_c(1S)}$, spreads over a range of $40-90\textmd{ MeV}$ \cite{Kwong1991,Eichten1994,Gershtein1995,Gupta1996,Fulcher1999,Ebert2003,Godfrey2004,Li2019}. The predictions of the excited states are even more scattered. First principle calculations are of cource essential to study the $B_c$ mesons with containable precision.
The first principle lattice QCD (lQCD) studies of the $B_c$ mesons are still lacking \cite{Allison:2004be,Gregory2010,Dowdall2012,Wurtz2015,Mathur2018}. A recent lQCD calculation predicts the $B_c$ meson masses with $J^P = 0^-,\, 1^-,\, 0^+,\, 1^+$ with high precision \cite{Mathur2018}, where $J$ is the angular momentum and $P$ is the $P$-parity ($C$ in the later text is the $C$-parity).

The Dyson-Schwinger equation and Bethe-Salpeter equation (DSBSE) approach, a nonperturbative quantum chromodynamic (QCD) approach solving QCD via the basic degree of freedom, i.e. the quarks and the gluons, is complementary to lQCD. In this approach, a meson corresponds to a pole in the quark-antiquark scattering kernel \cite{Itzykson1980}, and the amplitude is solved by the Bethe-Salpeter equation (BSE) \cite{Salpeter1951}. The propagators of quarks and gluons in the scattering kernel are then solved by the Dyson-Schwinger equations (DSE) \cite{Dyson1949,Schwinger1951}.  However, the DSEs are an infinite tower of coupled equations and no exact solution has been gained. A practical way to solve the problem of hadron spectrum is building a quark-gluon-vertex, constructing a scattering kernel, the gluon propagator being modelled by an effective interation, then the BSE of the hadron amplitude and DSE of the quark propagator could be solved consistently.
While the forms of the quark-gluon-vertex and the scattering kernel have been investigated \cite{Chang2009}, the most widely used and technically simple one is the rainbow ladder (RL) approximation. This scheme has gained many phenomenological succsess in the hadron spectrum, the form factors, the distribution functions, etc \cite{Roberts1994,Alkofer2001,Maris2003}.
The RL approximation is perfect for the light pseudoscalar mesons \cite{Maris1997,Maris1998} and fairly well for the light vector mesons \cite{Maris1999}. A possible way to solve the light meson with orbit angular moment $L\geq 1$ is given in Ref. \cite{Chang2009}.
The mass spectra of the charmonium and bottomonium with angular momentum $J<3$ have been investigated in the RL approximation, with relative errors a few percents \cite{Fischer2015,Hilger2015,Chen2017}. However, the original RL approximation fails to describe the heavy-light mesons due to the lacking of the flavor asymmetry. In Ref. \cite{Chen2019} we build an interaction model, which takes into account the quark flavor dependence of the effective quark-antiquark interaction properly and preserves the axial-vector Ward Takahashi identity excellently. Our model not only gives a successful unified description of the light, heavy-light and heavy pseudoscalar and vector mesons, but also reduces the error of the $B_c(1S)$ meson mass from $110\textmd{ MeV}$ \cite{Qin2018} to about $15\textmd{ MeV}$. In Ref. \cite{Chang2019} our model is applied to the radial excited $B_c(2S)$ and $B_c^*(2S)$ mesons. The mass spectrum is reasonable and we predict the decay constants of the radial excited $B_c$ mesons for the first time.

In this article, we extend our study to $B_c$ mesons with $J^P = 0^-, 1^-, 0^+, 1^+$ and $2^+$. In section \ref{model} we give a complete introduction of our model and the formulas. The results are analyzed in section \ref{results}. We summarize our investigations in section \ref{summary}.

\section{Our model}\label{model}

In this Poincar$\acute{\text{e}}$ covariant framework the quark propagator is solved by the Gap equation \cite{Dyson1949,Schwinger1951,Itzykson1980}
\begin{eqnarray}\nonumber
 S_f^{-1}(k) &=& Z_2 (i\gamma\cdot k + Z_m m_f) \\\label{eq:DSE0}
&&+ \frac{4}{3}\bar{g}^2 Z_1 \int^\Lambda_{d q} D_{\mu\nu}(l)\gamma_\mu S_{f}(q)\Gamma^f_\nu(k,q),
\end{eqnarray}
where $S_{f}$ is the quark propagator. $Z_1$, $Z_2$, $Z_m$ are the renormalisation constants of the quark-gluon-vertex, the quark field and the quark mass respectively.  $D_{\mu\nu}$ is the gluon propagator, $\Gamma^f_\nu$ the quark-gluon-vertex. 
$m_f$ is the current quark mass, $\bar{g}$ the coupling constant.
$f=\{u,d,s,c,b,t\}$ represents the quark flavor.
$l=k-q$, with $l$ the momentum of the gluon, $k$ and $q$ the momentum of outer and inner quark.
$\int^\Lambda_{d q}=\int ^{\Lambda} d^{4} q/(2\pi)^{4}$ stands for a Poincar$\acute{\text{e}}$ invariant regularized integration, with $\Lambda$ the regularization scale.
The Bethe-Salpeter amplitude (BSA) of the meson is solved by the BSE      \cite{Salpeter1951,Itzykson1980},
\begin{equation}\label{eq:BSE0}
  \big{[} \Gamma^{fg}(k;P)  \big{]}^{\alpha}_{\beta}  =   \int^\Lambda_{d q} \big{[} K^{fg}(k,q;P) \big{]}^{\alpha\delta}_{\sigma\beta} \big{[} \chi^{fg}(q;P)  \big{]}^{\sigma}_{\delta} ,
\end{equation}
where $\Gamma^{fg}$ is the BSA, $ K^{fg}$ the quark-antiquark scattering kernel. $k$ and $P$ are the relative and the total momentum of the meson, with $P^2 = -M^2$ and $M$ the meson mass. $\chi^{fg}(q;P) = S_{f}(q_{+}) \Gamma^{fg}(q;P) S_{g}(q_{-})$, where $\chi^{fg}$ is the wave function,
 and $q_{+} = q + \iota P/2$, $q_{-} = q - (1-\iota) P/2$.
 $\iota$ is the partitioning parameter describing the momentum partition between the quark and antiquark and dosen't affect the physical observables.
$\alpha$, $\beta$, $\sigma$ and $\delta$ are the Dirac indexes.

The RL approximation is making the following replacement in Eq.(\ref{eq:DSE0}) and Eq.(\ref{eq:BSE0}) \cite{Chen2019},
\begin{eqnarray}\label{eq:quarkRL}
&& \bar{g}^2 Z_1 D_{\mu\nu}(l) \Gamma^f_\nu(k,q) \to [Z_{2}]^{2} \tilde{D}^{f}_{\mu\nu}(l) \gamma_\nu, \\\label{eq:mesonRL}
&& \big{[} K^{fg}(k,q;P) \big{]}^{\alpha\delta}_{\sigma\beta} \to -\frac{4}{3}[Z_{2}]^{2} \tilde{D}^{fg}_{\mu\nu}(l) [\gamma_{\mu}^{}]^{\alpha}_\sigma [\gamma_{\nu}]^\delta_\beta,
\end{eqnarray}
where $\tilde{D}^{fg}_{\mu\nu}(l) = \left(\delta_{\mu\nu}-\frac{l_{\mu}l_{\nu}}{l^{2}}\right)\mathcal{G}^{fg}(l^2)$ and $\tilde{D}^{f}_{\mu\nu}(l) = \tilde{D}^{ff}_{\mu\nu}(l)$ are the effective quark-antiquark interactions.
The dressed function $\mathcal{G}^{fg}(s)$ is composed of a flavor dependent infrared part and a flavor independent ultraviolet part \cite{Chen2019},
\begin{eqnarray}\label{eq:gluonfmodel}
  \mathcal{G}^{fg}(s) 	&=& \mathcal{G}^{fg}_{IR}(s) + \mathcal{G}_{UV}(s),\\\label{eq:gluonfInfrared}
  \mathcal{G}^{fg}_{IR}(s) 	&=& 8\pi^2\frac{D_f}{\omega_f^2}\frac{D_g}{\omega_g^2} e^{-s/(\omega_f\omega_g)},\\\label{eq:gluonUltraviolet}
  \mathcal{G}_{UV}(s) 	&=& \frac{8\pi^{2} \gamma_{m}^{} \mathcal{F}(s)}{\text{ln}[\tau+(1+s/\Lambda^{2}_{QCD})^2]},
\end{eqnarray}
where $s=l^2$. $\mathcal{G}^{fg}_{IR}(s)$ is the infrared interaction responsible for dynamical chiral symmetry breaking (DCSB), with $(D_f^2\omega_f)^{1/3}$ expressing the interaction strength and $\omega_f$ the interaction width in the momentum space. The gaussian form is used as it enables the natural extraction of a monotonic running-coupling and gluon mass \cite{Qin2011}. $f$ and $g$ label the quark flavors. The equality in $f$ and $g$ means that the quark and antiquark contribute equally to the interaction strength and width.
$\mathcal{G}_{UV}(s)$ keeps the one-loop perturbative QCD limit in the ultraviolet. As we are dealing with 5 active quarks, $\mathcal{G}_{UV}(s)$ is independent of the quark flavors.
$\mathcal{F}(s)=[1 - \exp(-s^2/[4m_{t}^{4}])]/s$, $\gamma_{m}^{}=12/(33-2N_{f})$, with $m_{t}=1.0 \textmd{ GeV}\,$, $\tau=e^{10} - 1$, $N_f=5$, and $\Lambda_{\text{QCD}}=0.21 \textmd{ GeV}\,$.
The values of $m_{t}$ and $\tau$ are chosen so that $\mathcal{G}_{UV}(s)$ is well suppressed in the infrared and the dressed function $\mathcal{G}_{IR}^{fg}(s)$ is qualitatively right in the limit $m_f \to \infty$ or $m_g \to \infty$.

The paremeters $D_{f}$ and $\omega_{f}$ that express the flavor dependent quark-antiquark interaction are fitted by the physical observables. See Ref. \cite{Chen2019} for the detail.
With the parameters well fitted, the axial-vector Ward-Takahashi identity (av-WTI) is perfectly satisfied \cite{Chen2019}, which guarantees the ground state pseudoscalar mesons as Goldstone bosons of DCSB. Our model gives a successful and unified description of the light, heavy and heavy-light ground pseudoscalar and vector mesons \cite{Chen2019}.

\section{Results}\label{results}

Three sets of parameters of the charm and bottom system corresponding to the varying of the interaction width are listed in Table~\ref{tab:parametersBC}.
The current quark masses, defined in Ref. \cite{Chang2019}, are $\check{m}_{c}=1.31 \,\textmd{GeV}$ and $\check{m}_{b}=4.27 \, \textmd{GeV}$.
\begin{table}[t]
\vspace*{-5mm}
\caption{\label{tab:parametersBC} Three sets of the parameters $\omega_f$ and $D_f$ (in GeV) of the charm and bottom system \cite{Chen2019}.}
\begin{tabular}{c|c|c|c|c|c|c}
\hline
\multirow{2}{*}{flavor}&\multicolumn{2}{c|}{Para-1}&\multicolumn{2}{|c|}{Para-2}&\multicolumn{2}{|c}{Para-3}\\
\cline{2-7}
&\; $\omega_f $\; &\; $D_f^2$ \;&\; $\omega_f $\; &\; $D_f^2$ \;	 &\; $w_f $\; &\; $D_f^2$ \;\\ [0.5mm]
\hline
$c$		& 0.690 & 0.645 			& 0.730 & 0.599 			& 0.760 & 0.570 \\
$b$		& 0.722 & 0.258 			& 0.766 & 0.241 			& 0.792 & 0.231 \\
\hline
\end{tabular}
\end{table}
%

The masses of the charmonium are listed in Tab. \ref{tab:massccbar}.
$M^{\textmd{RL}}_{c\bar{c}}$ is our RL approximation result. $M^{\textmd{expt.}}_{c\bar{c}}$ is the experiment value \cite{Tanabashi2018}. $\Delta M^{\textmd{RL}}_{c\bar{c}}$ is the deviation of the RL results from the experiment value,
\begin{equation}\label{eq:deltaMccbar}
 \Delta M^{\textmd{RL}}_{c\bar{c}} = M^{\textmd{RL}}_{c\bar{c}} - M^{\textmd{expt.}}_{c\bar{c}}.
\end{equation}
The mass of the pseudoscalar ($J^{PC} = 0^{-+}$) meson is used to fit the parameters, so there is no deviation for it. For all the other mesons, the deviations are less than $3.5\%$. Three sets of parameters are used in our calculation. The uncertainties due to the varying of the parameters are very small. The absolute uncertainty is $4\textmd{ MeV}$ for the vector ($J^{PC} = 1^{--}$) meson, and $5\textmd{ MeV}$ for the scalar meson ($J^{PC} = 0^{++}$). For other mesons ($J^{PC} = 1^{+-},\, 1^{++},\,2^{++}$) the absolute uncertainties are less than $26\textmd{ MeV}$, with the relative uncertainties less than $0.8\%$.

\begin{table}[t!]
\centering
\caption{\label{tab:massccbar} Masses (in MeV) of the charmonium with $J^{PC}\,=\,0^{-+},\,1^{--},\,0^{++},\,1^{+-},\,1^{++},\,2^{++}$, the normal states in the quark model. $M^{\textmd{RL}}_{c\bar{c}}$ is our RL approximation result. $M^{\textmd{expt.}}_{c\bar{c}}$ is the experiment value \cite{Tanabashi2018}. $\Delta M^{\textmd{RL}}_{c\bar{c}} = M^{\textmd{RL}}_{c\bar{c}} - M^{\textmd{expt.}}_{c\bar{c}}$ is the deviation of our results from the experiment value. Three sets of parameters in Tab. \ref{tab:parametersBC} are used in our calculation. }
\begin{tabular}{c|c|c|c|c|c|c|c}
\hline    
\multicolumn{2}{c|}{$J^{\textmd{PC}}$} & $0^{-+}$ & $1^{--}$ & $0^{++}$ & $1^{+-}$ & $1^{++}$ & $2^{++}$\\
\hline
\multirow{3}{*}{$M^{\textmd{RL}}_{c\bar{c}}$} 
& Para-1 & 2984 & 3134 & 3327 & 3400 & 3417 & 3497 \\
& Para-2 & 2984 & 3132 & 3331 & 3416 & 3426 & 3511 \\
& Para-3 & 2984 & 3130 & 3332 & 3426 & 3431 & 3518 \\
\hline
\multirow{3}{*}{$\Delta M^{\textmd{RL}}_{c\bar{c}}$}
& Para-1 & 0     & 37 & -88& -125& -94& -59\\
& Para-2 & 0     & 35 & -84& -109& -85& -45\\
& Para-3 & 0     & 33 & -83& -99& -80& -38\\
\hline
\multicolumn{2}{c|}{$M^{\textmd{expt.}}_{c\bar{c}}$}
& 2984 & 3097 & 3415 & 3525 & 3511 & 3556\\
\hline
\end{tabular}
\end{table}

\begin{table}[t!]
\centering
\caption{\label{tab:massbbbar} Masses (in MeV) of the bottomonium. The meanings of the quantities are the same as in Tab. \ref{tab:massccbar}.}
\begin{tabular}{c|c|c|c|c|c|c|c}
\hline    
\multicolumn{2}{c|}{$J^{\textmd{PC}}$} & $0^{-+}$ & $1^{--}$ & $0^{++}$ & $1^{+-}$ & $1^{++}$ & $2^{++}$\\
\hline
\multirow{3}{*}{$M^{\textmd{RL}}_{b\bar{b}}$} 
& Para-1 & 9399 & 9453 & 9754 & 9793 & 9788 & 9820\\
& Para-2 & 9399 & 9453 & 9762 & 9805 & 9799 & 9833\\
& Para-3 & 9399 & 9453 & 9765 & 9810 & 9804 & 9835\\
\hline
\multirow{3}{*}{$\Delta M^{\textmd{RL}}_{b\bar{b}}$}
& Para-1 & 0     & -7& -105& -106& -106& -92\\
& Para-2 & 0     & -7& -97& -94& -94& -79\\
& Para-3 & 0     & -7& -94& -89& -89& -77\\
\hline
\multicolumn{2}{c|}{$M^{\textmd{expt.}}_{b\bar{b}}$}
& 9399 & 9460 & 9859 & 9899 & 9893 & 9912\\
\hline
\end{tabular}
\end{table}

The masses of the bottomonium are listed in Tab. \ref{tab:massbbbar}. $M^{\textmd{RL}}_{b\bar{b}}$ and $M^{\textmd{expt.}}_{b\bar{b}}$ are our RL approximation result and the experiment value \cite{Tanabashi2018} respectively. $\Delta M^{\textmd{RL}}_{b\bar{b}}$ is the deviation, defined by
\begin{equation}\label{eq:deltaMbbbar}
 \Delta M^{\textmd{RL}}_{b\bar{b}} = M^{\textmd{RL}}_{b\bar{b}} - M^{\textmd{expt.}}_{b\bar{b}}.
\end{equation}
There is no mass deviation for the pseudoscalar meson as it is used to fit the parameters. For all the other mesons, the deviations are less than $1.1\%$. The uncertainty due to the varying of the parameters is very small for the vector meson (less than $1\textmd{ MeV}$). For other mesons ($J^{PC} = 0^{++},\, 1^{+-},\, 1^{++},\,2^{++}$) the absolute uncertainties are less than $17\textmd{ MeV}$, with the relative uncertainties less than $0.2\%$.

The results in Tab. \ref{tab:massccbar} and Tab. \ref{tab:massbbbar} have two-sided meanings. On one hand, the RL approximation is reasonble for the charmonium and bottomonium system. The mass deviations are less than $3.5\%$ for the charmonium and less than $1.1\%$ for the bottomonium. However, the absolute errors do not decrease for the P-wave states, indicating that in the RL approximation some interactions are still lacking even for the heavy system. On the other hand, the results are stable when the parameters change. The uncertainties due to the varying of the parameters are small, less than $0.8\%$ for the charmonium and less than $0.2\%$ for the bottomonium. So the errors (defined by Eq.(\ref{eq:deltaMccbar}) and Eq.(\ref{eq:deltaMbbbar})) of the RL approximation could be estimated quantitatively.
The interaction, Eq.(\ref{eq:gluonfmodel}) $\sim$ Eq.(\ref{eq:gluonUltraviolet}), expresses the flavor dependence properly, so that the errors are of the same order for
both the open flavor mesons and the $q\bar{q}$ mesons \cite{Chen2019}.
As the masses of the $B_c$ mesons are approximately $M_{c\bar{b},J^P} \approx (M_{c\bar{c},J^P} + M_{b\bar{b},J^P})/2$, the mass errors of the $B_c$ mesons are assumed to be:
\begin{equation}\label{eq:errorMbc}
 \Delta M^{\textmd{RL}}_{c\bar{b},J^P} = \frac{1}{N_C}\sum_{C} (\Delta M^{\textmd{RL}}_{c\bar{c},J^{PC}} + \Delta M^{\textmd{RL}}_{b\bar{b},J^{PC}})/2,
\end{equation}
with $J^{PC}$ the normal states, $N_C$ the number of the normal states.
For example, if $J^P = 0^-$, then the corresponding normal $c\bar{c}$ or $b\bar{b}$ state is $J^{PC} = 0^{-+}$, and $N_C = 1$. $\Delta M^{\textmd{RL}}_{c\bar{b},0^-} = (\Delta M^{\textmd{RL}}_{c\bar{c},0^{-+}} + \Delta M^{\textmd{RL}}_{b\bar{b},0^{-+}})/2$. It's the same case for the $J^P = 1^-,\,0^+,\,2^+$ mesons. If $J^P = 1^+$, then both $J^{PC} = 1^{++}$ and $1^{+-}$ are the normal states and $N_C = 2$. The $J^P = 1^+$ $B_c$ mesons are the mixing states of the $^1 P_1$ state and $^3 P_1$ state \cite{Li2019}. We do not bother to investigate the mixing. The mass errors of the $J^P = 1^+$ mesons are estimated to be 
$
 \Delta M^{\textmd{RL}}_{c\bar{b},1^+} = (\Delta M^{\textmd{RL}}_{c\bar{c},1^{++}} + \Delta M^{\textmd{RL}}_{c\bar{c},1^{+-}} + \Delta M^{\textmd{RL}}_{b\bar{b},1^{++}} + \Delta M^{\textmd{RL}}_{b\bar{b},1^{+-}})/4$,
i.e., averaging the $C$-parity. In Tab. \ref{tab:massccbar} and  Tab. \ref{tab:massbbbar}, the difference of $\Delta M^{\textmd{RL}}_{c\bar{c}}$ with $J^{PC}=1^{++}$ and $J^{PC}=1^{+-}$ are less than $31\textmd{ MeV}$, and there is no difference for $\Delta M^{\textmd{RL}}_{b\bar{b}}$ for these two states. So taking the $C-$parity average leads no more than $8\textmd{ MeV}$ error for $\Delta M^{\textmd{RL}}_{c\bar{b},1^+}$. 

\begin{table}[t!]
\caption{\label{tab:massExcited}
The masses (in MeV) of the first radial excited states of the charm-bottom system with $J^{P}=0^{-}$ (cited from Ref. \cite{Chang2019}).
The experiment data for $M_{\eta_{c}(2S)}$ and $M_{\eta_{b}(2S)}$ are taken from Ref.~\cite{Tanabashi2018},
and that for $M_{B^+_{c}(2S)}$ is taken from Ref.~\cite{Aaij2019}.}
\begin{tabular}{c|c|c|c}
\hline
state	&$\eta_{c}(2S)$	&$B^+_{c}(2S)$	&$\eta_{b}(2S)$	\\
\hline
RL	    &$3606$		&$6813$	&$9915$		\\
expt.	&$3638$		&$6872$	&$9999$		\\
$\Delta M^{\textmd{RL}}$&-32&-59&-84\\
\hline
\end{tabular}
\end{table}

The relation Eq.(\ref{eq:errorMbc}) is also supported by the masses of the radial excited states \cite{Chang2019}. The center values of the excited state masses are cited in Tab. \ref{tab:massExcited}. $(\Delta M^{\textmd{RL}}_{\eta_{c}(2S)} + \Delta M^{\textmd{RL}}_{\eta_{b}(2S)})/2 = -58\textmd{ MeV}$, while $\Delta M^{\textmd{RL}}_{B^+_{c}(2S)} = -59\textmd{ MeV}$. The direct calculation of the $B_c$ meson masses in the RL approximation and the mass errors due to the RL approximation are listed in Tab. \ref{tab:masscbRL}. The modified mass is defined by
\begin{equation}\label{eq:modifiedMass}
 \bar{M}^{\textmd{RL}}_{c\bar{b}} = M^{\textmd{RL}}_{c\bar{b}} - \Delta M^{\textmd{RL}}_{c\bar{b}}.
\end{equation}
With the errors from the RL approximation subduced, $\bar{M}^{\textmd{RL}}_{c\bar{b}}$ are our prediction for the $B_c$ mesons. There are two kinds of other errors for all the $B_c$ mesons, which are estimated as following.
$M^{\textmd{RL}}_{c\bar{c},0^{-+}}$ and $M^{\textmd{RL}}_{b\bar{b},0^{-+}}$ are used to fit the parameters, so the error of $M^{\textmd{RL}}_{c\bar{b},0^{-}}$ is totally due to the interaction pattern, Eq.(\ref{eq:gluonfmodel}) $\sim$ Eq.(\ref{eq:gluonUltraviolet}). The error due to this interaction pattern, the first error, is about $15\textmd{ MeV}$. The results vary by a few MeVs as the parameters change, which is the second error. These errors due to the varying of the paramters are much smaller than those of the charmonium and the bottomonium. The uncertainties of the parameters cancel by using Eq.(\ref{eq:errorMbc}), i.e., inferring the errors of the $B_c$ mesons as the intermidiate of the charmonium and the bottomonium. For the $J^P = 1^+$ $B_c$ mesons, there is a third error due to the $C-$parity average in Eq.(\ref{eq:errorMbc}), which is about $8\textmd{ MeV}$.

\begin{table}[t!]
\centering
\caption{\label{tab:masscbRL} Masses of the $B_c$ Mesons (in MeV). $M^{\textmd{RL}}_{c\bar{b}}$ is the direct RL result. $\Delta M^{\textmd{RL}}_{c\bar{b}}$ is the error of the RL approximation defined by Eq.(\ref{eq:errorMbc}). $\bar{M}^{\textmd{RL}}_{c\bar{b}}$ is the modified mass, defined by Eq.(\ref{eq:modifiedMass}).}
\begin{tabular}{c|c|c|c|c|c|c|c}
\hline    
\multicolumn{2}{c|}{$J^{\textmd{P}}$} & $0^{-}$ & $1^{-}$ & $0^{+}$ & $1_1^{+}$ & $1_2^{+}$ & $2^{+}$\\
\hline
\multirow{3}{*}{$M^{\textmd{RL}}_{c\bar{b}}$} 
& Para-1 & 6293 & 6360 & 6608 & 6642 & 6677 & 6721\\
& Para-2 & 6290 & 6357 & 6612 & 6649 & 6686 & 6731\\
& Para-3 & 6287 & 6354 & 6612 & 6651 & 6688 & 6733\\
\hline
\multirow{3}{*}{$\Delta M^{\textmd{RL}}_{c\bar{b}}$} 
& Para-1 & 0     & 15 & -97& -108 & -108 & -75\\
& Para-2 & 0     & 14 & -91& -96 & -96 & -62\\
& Para-3 & 0     & 13 & -89& -89 & -89 & -57\\
\hline
\multirow{3}{*}{$\bar{M}^{\textmd{RL}}_{c\bar{b}}$} 
& Para-1 & 6293 & 6345 & 6705 & 6750 & 6785 & 6796\\
& Para-2 & 6290 & 6343 & 6703 & 6745 & 6782 & 6793\\
& Para-3 & 6287 & 6341 & 6701 & 6740 & 6777 & 6790\\
\hline
\end{tabular}
\end{table}

\begin{table}[t!]
\caption{\label{tab:masscball} Masses of the $B_c$ Mesons (in MeV).  $\bar{M}^{\textmd{RL}}_{c\bar{b}}$ is our prediction. The first error is due to the interaction pattern Eq.(\ref{eq:gluonfmodel}) $\sim$ Eq.(\ref{eq:gluonUltraviolet}). The second error is due to the varying of the parameters. For $J^{P} = 1^+$ mesons, the third error is due to the $C-$parity average in Eq.(\ref{eq:errorMbc}). $M^{\textmd{QM}}_{c\bar{b}}$ is the quark model result \cite{Li2019}, and the underlined ones are the input values. $M^{\textmd{LQCD}}_{c\bar{b}}$ is the lQCD prediction \cite{Mathur2018}. $M^{\textmd{expt.}}_{c\bar{b}}$ is the experiment value \cite{Tanabashi2018}.}
\begin{tabular}{c|c|c|c|c}
\hline    
$J^{\textmd{P}}$  & $\bar{M}^{\textmd{RL}}_{c\bar{b}}$	&	$M^{\textmd{QM}}_{c\bar{b}}$ &	$M^{\textmd{LQCD}}_{c\bar{b}}$ & $M^{\textmd{expt.}}_{c\bar{b}}$ \\
\hline
$0^{-}$  & 6290(15)(3)  & \underline{6271} & 6276(7)   & 6275(1) \\
$1^{-}$  & 6343(15)(2)  & \underline{6326} & 6331(7)   & -- \\
$0^{+}$	 & 6703(15)(2)  & 6714 & 6712(19)  & --  \\
$1_1^{+}$& 6745(15)(5)(8)  & 6757 & 6736(18)  & -- \\
$1_2^{+}$& 6781(15)(4)(8)  & 6776 & --         & -- \\
$2^{+}$	 & 6793(15)(3)  & 6787 & --         & -- \\
\hline
\end{tabular}
\end{table}

Our predictions of the $B_c$ mesons and the estimated errors are listed in the second column of Tab.\ref{tab:masscball}. Hitherto the only experiment data for the $B_c$ spectrum is the pseudoscalar meson mass, as the production rate of the $B_c$ mesons is much lower than those of the charmonium and the bottomonium. Our results are consistent with the recent lQCD predictions (with $J^P = 0^-,\,1^-,\,0^+,\,1^+$), which are listed in the forth column. Our results are also consistent with the quark model predictions. The quark model predictions from Ref. \cite{Li2019} are listed in the third column, where the masses with $J^P =0^-,\,1^-$ are the input values and others are the outputs. The mass splitting of the $1S$ state of our result is $M_{B_c(1^-)} - M_{B_c(0^-)} = 53\textmd{ MeV}$, consistent with the lQCD result $55\textmd{ MeV}$. The mass splittings of the $1P$ states of our results are $(M_{B_c(1^+_1)} + M_{B_c(1^+_2)})/2 - M_{B_c(0^+)} = 60\textmd{ MeV}$ and $M_{B_c(2^+)} - M_{B_c(0^+)} = 90\textmd{ MeV}$, while the quark model results are $53\textmd{ MeV}$ and $73\textmd{ MeV}$ respectively. Our results of the $B_c$ meson masses also have two-sided meanings. On one hand, we predict the masses of the $B_c$ mesons (with $J^P = 0^-,\,1^-,\,0^+,\,1^+,\,2^+$), providing a significant guide to the experimental search for the $B_c$ mesons. On the other hand, the consistency of our results with other predictions supports that the flavor dependent interaction pattern, Eq.(\ref{eq:gluonfmodel}) $\sim$ Eq.(\ref{eq:gluonUltraviolet}), is reasonable. This pattern leads an error about $15\textmd{ MeV}$ for the $B_c$ mesons.


\section{Summary}\label{summary}

In this paper, we predict the masses of the $B_c$ mesons with $J^P = 0^-,\,1^-,\,0^+,\,1^+,\,2^+$ using a flavor dependent interaction pattern via the Dyson-Schwinger equation and Bethe-Salpeter equation approach. This interaction pattern, composed of a flavor dependent infrared part and a flavor independent ultraviolet part, gives an unified successful description of the pseudoscalar and vector light, heavy-light and heavy mesons. This interaction pattern could also be applied to the radial excited heavy mesons. Herein we control the errors carefully. Besides the error from the RL approximation, two other kinds of errors are considered. One is the error from the interaction pattern, the other is the error from the varying of the parameters. For the $J^P = 1^+$ $B_c$ mesons, a third error due to the $C-$parity average is also considered. With the errors from the RL approximation subduced, our predictions are consistent with the lQCD and quark model results. Our results have two-sided meanings. On one hand, we predict the $B_c$ meson masses with errors well controlled, providing a significant guide to the experiment search.  On the other hand, the excellent consistency of our predictions with the results from other approaches also supports strongly that the flavor dependent interaction pattern, Eq.(\ref{eq:gluonfmodel}) $\sim$ Eq.(\ref{eq:gluonUltraviolet}), is reasonable.

\section*{Acknowledgments}
We acknowledge helpful conversations with Xian-Hui Zhong, Qi Li, Ming-Sheng Liu and Long-Cheng Gui. Chen thanks Xian-Hui Zhong for the surport and the encourage in publishing this paper. This work is supported by: the National Natural Science Foundation of China under contracts No. 11947108, the Chinese Government Thousand Talents Plan for Young Professionals and the National Natural Science Foundation of China under contracts No. 11435001, and No. 11775041, the National Key Basic Research Program of China under contract No. 2015CB856900. 

\bibliographystyle{unsrt}
\bibliography{BcAll}

\end{document}